# Describing NMR chemical exchange by effective phase diffusion approach

Guoxing Lin*

Carlson School of Chemistry and Biochemistry, Clark University, Worcester, MA 01610, USA

*Email: glin@clarku.edu

**Abstract**

This paper proposes an effective phase diffusion method to analyze chemical exchange in nuclear magnetic resonance (NMR). The chemical exchange involves spin jumps around different sites where the spin angular frequencies vary, which is a random phase walk viewed from the rotating frame reference. Such a random walk in phase space can be treated by the effective phase diffusion method. Both the coupled and uncoupled normal and fractional phase diffusions are considered. Based on the phase diffusion results, the line shape of NMR exchange spectrum can be obtained. By comparing these theoretical results with the conventional theory, this phase diffusion approach works for fast exchange, ranging from slightly faster than intermediate exchange to very fast exchange. For normal diffusion models, the theoretically predicted curves agree with those predicted from traditional models in the literature, and the characteristic exchange time obtained from phase diffusion with a fixed jump time is the same as that obtained from the conventional model. However, the phase diffusion with a monoexponential time distribution gives a relatively shorter characteristic exchange time constant, which is half of that obtained from the traditional model. Additionally, the fractional diffusion obtains a significantly different line shape than that predicted based on normal diffusion.

## 1. Introduction

Molecule exchange involves spin jumping around different environmental sites due to diffusion or changes in conformational or chemical states [1,2]. Nuclear magnetic resonance (NMR) is a powerful technique to detect exchange processes in biological and polymer systems at the atomic level [1,2,3,4]. The exchange process can be investigated by NMR via the signal changes resulting from various factors, such as the diffusion coefficient [5], relaxation rate [6,7], and angular frequency of precession [1,2,8,9]. The angular frequency of the spin precession changes at different exchange sites, which results in observable line shape changes in the NMR spectrum. Analysis of the line shape change is one of the fundamental NMR tools to investigate exchange dynamics and will be the focus of this paper. Although the NMR chemical exchange has been an established tool, theoretical developments are still needed to understand the chemical exchange NMR better, particularly for complex systems.

In exchange, the characteristic exchange time, an essential dynamic parameter, could follow a complicated distribution. The exchange time is a principal factor affecting the NMR line shape of chemical exchange. Many theoretical models have been developed to analyze chemical exchanges in NMR [1,2,5,8,10]. The two-site exchange model based on the modified-Bloch equation successfully interprets many NMR lines shapes [1,8], where the jump time in the exchange equation is a fixed constant. However, in a real system, the exchange time could follow a distribution such as the exponential function in the Gaussian exchange model reported in [11]. A complex exchange distribution could also exist in complicated conformational change or diffusion-induced exchange, such as Xenon diffusing in the heterogeneous system [12]. In a complex system, a monoexponential time distribution may not be sufficient to explain the dynamics behavior. For complicated systems, the time distribution function could be the Mittag-Leffler function (MLF) $E_\alpha\left(-\left(\frac{t}{\tau}\right)^\alpha\right)$ [13,14], or a stretched exponential function (SEF) $\exp\left(-\left(\frac{t}{\tau}\right)^\alpha\right)$, where $\alpha$ is the time-fractional derivative order, and $\tau$ is the characteristic time. The MLF $E_\alpha(-t^\alpha) = \sum_{n=0}^\infty \frac{(-t^\alpha)^n}{\Gamma(n\alpha+1)}$, can be reduced to a SEF $\exp\left(-\frac{t^\alpha}{\Gamma(1+\alpha)}\right)$ when $t$ is small. The SEF $\exp\left(-\left(\frac{t}{\tau}\right)^\alpha\right)$ is the same as the Kohlrausch-Williams-Watts (KWW) function [15,16,17,18], a well-known time correlation function in macromolecular systems. The SEF has been used in explaining pulsed-field gradient (PFG) or relaxation NMR [19,20] and magnetic resonance imaging (MRI) [21,22]. The Mittag-Leffler function-based distribution is heavy-tailed. Mittag Leffler function has been employed to analyze anomalous NMR dynamics processes such as PFG anomalous diffusion [19,23,24,25] and anomalous NMR relaxation [26,27,28,29]. Currently, the chemical exchange theories of NMR are still difficult to handle the chemical exchange with these complex distributions.

A phase diffusion method is proposed in this paper to explain the NMR chemical exchange. Most current methods are real space approaches, such as the modified Bloch exchange equations [1, 2], while the Gaussian exchange model [11] is a phase space method based on evaluating the accumulated phase variance for the random phase process during the exchange process. Because the spin phase in the chemical exchange undergoes a random walk in the rotating frame reference, an effective phase diffusion method will be proposed in this paper to analyze the exchange. Effective phase diffusion has been applied to analyze PFG diffusion and NMR relaxation [19,26]. It has certain advantages over the traditional methods: It can provide the exact phase distribution that may be difficult to obtain by conventional real space theoretical method; additionally, the NMR signal can be directly obtained from vector sum by Fourier transform in phase space, which makes the analysis intuitive and often simplifies the solving process; furthermore, the phase diffusion method could be straightforwardly applied to anomalous dynamics process based on fractional calculus [30, 31, 32,33,34]. Fractional diffusion arises from diffusion waiting time distribution behaving asymptotically to a power law, or from its jump length distribution following a power law [35]. Fractional calculus could help researchers analyze complex systems and provide more detailed information [36], which may not be interpretable by the ordinary method [24]. The fractional model could provide a better or more convenient fitting to experimental data than the traditional model. In Ref. [26,29], it shows that the fractional expression uses fewer parameters in fitting the experimental spin-lattice relaxation data. In Ref. [24], the apparent diffusion coefficient based on restricted fractional diffusion provides improved fitting to experimental data than that by the traditional model. Both the normal and fractional phase diffusions are considered, where the exchange time can be a simple constant or follow a certain type of distribution.

Additionally, each phase jump length is proportional to the jump time and the angular frequency. Because both the jump time and angular frequency fluctuate and obey certain types of distributions, the distribution of phase jump length could be either strongly or weakly correlated to the jump time distribution. The phase walk with weak phase time correlation can be treated by the uncoupled diffusion, while the strong correlation may require coupled diffusion model [35,36]. The traditional uncoupled diffusion has been successful in explaining many transport phenomena. However, it is insufficient to account for the divergence of the second moment of Levy flight processes [36], where a coupled diffusion is needed. From the coupled diffusion theory, the cooping between jump time and jump length significantly changes the variance of the random walk's jump length [36], and thus the diffusion coefficient, which could profoundly impact the phase diffusion outcome in the NMR line shape.

The rest of the paper is organized as follows. Section 2.1 treats the simplest normal diffusion with a fixed jump time: The obtained exchange time agrees well with the traditional two-site exchange. While the phase diffusion with random walk waiting time distribution is presented in Section 2.2: Firstly, the general expressions for phase evolution in chemical exchange are derived in Section 2.2.1; secondly, the normal diffusion is presented in Section 2.2.2, with both uncoupled and coupled diffusion, and it is found that the exchange time constant for diffusion with monoexponential time distribution is two-time faster than that of the traditional model and the fixed jump time diffusion result; thirdly, the fractional diffusion with MLF based jump time distribution is derived in Section 2.2.3, where the uncoupled diffusion is handled by time-fractional diffusion equation, and the coupled fractional diffusion is handled by coupled random walk model [36]. The results here give additional insights into the NMR chemical exchange, which could improve the analysis of NMR and magnetic resonance imaging (MRI) experiments, particularly in complicated systems.

## 2. Theory

The chemical exchange occurs when the spin jumps among different sites where the spin precession frequencies are different [3,4,8]. The precession frequencies of the spin moment are proportional to the intensity of the local magnetic field, which is affected by the surrounding electron cloud and nearby spin moments [4]. For simplicity, we consider only the basic exchange between two sites with equal populations [1,8] in this paper, neglecting the relaxation effect. The average precession angular frequencies for these two sites are arbitrarily set as $\omega_1$ and $\omega_2$ respectively, with $\omega_1 < \omega_2$ and
$$\Delta\omega = \omega_2 - \omega_1.$$

If the angular frequency of the rotating frame reference is set as $\frac{\omega_1+\omega_2}{2}$; the two sites have relative angular frequencies -$\omega_0$ and $\omega_0$, respectively, with $\omega_0 = \frac{\omega_2-\omega_1}{2}$.

The phase of spin undergoes chemical exchange during a time interval $\tau$ changes either by $\omega_0\tau$ or -$\omega_0\tau$ depending on the site. From the traditional exchange equations for chemical exchange, the spin always jumps to a site with a different angular frequency after a time interval $\tau$. Here, the choice for the next sites' frequency is assumed to be random, either the same or different. This assumption may be more realistic; for instance, after a time interval $\tau$, a spin may successfully jump to another site or return to the original site; or for the diffusion-induced exchange for Xenon spin in a heterogeneous system, the Xenon spin moves randomly to a similar environment with the same frequency or a different environment with a different frequency. The random walk process in phase space can be analyzed by phase diffusion [19,26]. Both the normal and fractional phase diffusion will be considered in the following.

*2.1 Simple normal diffusion with a fixed jump time*

If the jump time interval $\tau$ is a constant, the random phase jumps with average jump length $\Delta\phi$ equaling $-\omega_0\tau$ or $\omega_0\tau$. The effective phase diffusion constant $D_{\phi,s}$ for such a simple normal diffusion can be obtained by [23]

$$D_{\phi,s} = \frac{\langle(\Delta\phi)^2\rangle}{2\tau} = \frac{(\omega_0\tau)^2}{2\tau} = \frac{\omega_0^2}{2}\tau, \tag{1}$$

and the normal phase diffusion equation can be described by [19,26]

$$\frac{dP(\phi,t)}{dt} = D_{\phi,s}\Delta P(\phi,t), \tag{2}$$

where $\phi$ is the phase and $P(\phi,t)$ is the probability density function (PDF) of spin at time $t$ with $\phi$. The solution of Eq. (2) is [19]

$$P(\phi,t) = \frac{1}{\sqrt{4\pi D_{\phi,s}t}}\exp\left[-\frac{\phi^2}{4D_{\phi,s}t}\right]. \tag{3}$$

The total magnetization $M(t)$ is obtained by

$$M(t) = \int_{-\infty}^{\infty}d\phi\, e^{i\phi}P(\phi,t) = \exp(-D_{\phi,s}t), \tag{4}$$

which is a time-domain signal. By Fourier transform, we have the frequency domain signal

$$S(\omega) = \frac{D_{\phi,s}}{D_{\phi,s}^2 + \omega^2} = \frac{\frac{\omega_0^2}{2}\tau}{\left(\frac{\omega_0^2}{2}\tau\right)^2 + \omega^2}. \tag{5}$$

Note that both $\omega$ and $\omega_0$ are the angular frequencies in the rotating frame reference.

**2.2** *Diffusion with waiting time distribution*

*2.2.1 General expressions for phase evolution in chemical exchange*

A more realistic exchange time should follow a certain type of time distribution function $\varphi(t)$, which is often related to the time correlation function $G(t)$ by

$$\varphi(t) = -\frac{dG(t)}{dt}, \tag{6a}$$

and inversely

$$G(t) = 1 - \int_0^t \varphi(t')dt'. \tag{6b}$$

A commonly used simple time correlation function is the mono exponential distribution; in contrast, in a complicated system, it can be a Mittag Leffler function [29] or stretched exponential function such as the KWW function [15-18].

For a spin from the site with frequency $\omega_{i,0}$, its probability of starting jumps at a time $t'$ and acquiring phase $\omega_{i,0}t' + \phi$ at time $t$ is

$$P_{\omega_{i,0}}(\omega_{i,0}t' + \phi, t) = \varphi(t')P(\phi, t - t'), \tag{7}$$

where the phase change $\omega_{i,0}t'$ is obtained from time 0 to time $t'$ when the spin stays immobile at the site, and $P(\phi, t - t')$ is the phase PDF resulting from the diffusion or random walk in the phase space during $t - t'$. Summing all possible magnetization vectors with different phase $\omega_{i,0}t' + \phi$, at time $t$, the net magnetization $M_{\omega_{i,0}}(t', t)$ contributed from these spins beginning to jump randomly from time $t'$ is

$$M_{\omega_{i,0}}(t', t) = \int_{-\infty}^{\infty} d\phi \, e^{i\omega_{i,0}t' + \phi} P_{\omega_{i,0}}(\phi, t - t') = e^{i\omega_{i,0}t'} \varphi(t') p(k, t - t')|_{k=1}, \tag{8a}$$

where

$$p(k, t - t')|_{k=1} = \int_{-\infty}^{\infty} d\phi \, e^{ik\phi} P(\phi, t - t'). \tag{8b}$$

The magnetization from all diffusing spins in the systems at time t is

$$M_d(t) = \int_0^t dt' \sum_i p_i M_{\omega_{i,0}}(t', t), \tag{9a}$$

where $p_i$ is the population of spins at the $i^{th}$ site type. While, at time $t$, there are still spins that have not moved yet, and the magnetization from these still non-diffusing spins (or called residual spins), $M_{rs}(t)$, is

$$M_{rs}(t) = \sum_i p_i e^{i\omega_{i,0}t}(1 - \int_0^t \varphi(t')dt') = \sum_i p_i e^{i\omega_{i,0}t} G(t). \tag{9b}$$

The total magnetization $M(t)$ is the combination of $M_d(t)$ and magnetization from diffusing spins with the magnetization $M_{rs}(t)$ for those still non-diffusing spins (residual spins),

$$M(t) = M_d(t) + M_{res}(t) = \int_0^t dt' \sum_i p_i M_{\omega_{i,0}}(t', t) + \sum_i p_i e^{i\omega_{i,0}t} G(t). \tag{9c}$$

For simplicity, only exchange with two equal population sites will be considered here; let $\omega_{1,0} =$

$-\omega_0$, $\omega_{2,0} = \omega_0$, and the subindex $i$ will be dropped off throughout the rest of the paper. For a two-site system with equal populations $p_1 = p_2 = \frac{1}{2}$, from Eqs. (9a-b), we have

$$M_{rs}(t) = \frac{1}{2}[e^{-i\omega_0 t} + e^{i\omega_0 t}]G(t) = \cos(\omega_0 t)\, G(t), \tag{10a}$$

$$M_d(t) = \int_0^t dt' \frac{1}{2}[M_{-\omega_0}(t',t) + M_{\omega_0}(t',t)]$$

$$= \int_0^t dt' \frac{1}{2}\left[e^{-i\omega_0 t'} + e^{i\omega_0 t'}\right]\varphi(t')p(k,t-t')|_{k=1}$$

$$= \int_0^t dt'\, B(t')p(k,t-t')|_{k=1}, \tag{10b}$$

where

$$B(t') = \frac{1}{2}[e^{-i\omega_0 t'} + e^{i\omega_0 t'}]\varphi(t'). \tag{10c}$$

Eq. (10b) involves the convolution of $B(t')$ and $p(k,t-t')|_{k=1}$, thus in Laplace representation 34],

$$M_d(s) = B(s)p(k,s)|_{k=1}; \tag{11}$$

and the frequency domain signal can be obtained by the Fourier transform of $M(t)$:

$$S(\omega) = \int_0^\infty e^{i\omega t}M(t)dt = S_d(\omega) + S_{rs}(\omega), \tag{12a}$$

where

$$S_{rs}(\omega) = \int_0^\infty e^{i\omega t}M_{rs}(t)dt = \frac{1}{2}[G(\omega - \omega_0) + G(\omega + \omega_0)], \tag{12b}$$

$$S_d(\omega) = \int_0^\infty e^{i\omega t}M_d(t)dt = B(\omega)p(k,\omega)|_{k=1} \tag{12c}$$

In Eq. (12b), the $S_{rs}(\omega)$ is obtained straightforwardly from the Fourier cosine transform of the time correlation function. In the following, we will focus on how to obtain $M_d(t)$ and $S_d(\omega)$ for these diffusion spins in different diffusion scenarios.

2.2.2 *Normal diffusion with monoexponential distribution function*

Now, let us consider the case where the jump time follows a monoexponential distribution $\varphi(t)$ described by [35,36, 37]

$$\varphi(t) = \frac{1}{\tau}\exp\left(-\frac{t'}{\tau}\right), \tag{13a}$$

whose corresponding time correlation function is $G(t) = \exp\left(-\frac{t'}{\tau}\right)$. Laplace representation of $\varphi(t)$ is

$$\varphi(s) = \frac{1}{s\tau + 1}. \tag{13b}$$

Based on Eq. (12b), the frequency domain signal contributed from these residual spins are

$$S_{rs}(\omega) = \frac{1}{2}\left[\frac{\tau}{[\tau(\omega-\omega_0)]^2+1} + \frac{\tau}{[\tau(\omega+\omega_0)]^2+1}\right]. \tag{14}$$

According to Eq. (10c),

$$B(t) = \frac{1}{2}[e^{-i\omega_0 t'} + e^{i\omega_0 t'}]\frac{1}{\tau}\exp\left(-\frac{t'}{\tau}\right), \tag{15a}$$

whose Laplace representation is [34-36]

$$B(s) = \frac{1}{2}\left[\frac{1}{\tau(s-i\omega_0)+1} + \frac{1}{\tau(s+i\omega_0)+1}\right] \approx \frac{1}{1+\omega_0^2\tau^2+\tau s(1-\omega_0^2\tau^2)} = \frac{\frac{1}{1+\omega_0^2\tau^2}}{1+s\frac{\tau(1-\omega_0^2\tau^2)}{1+\omega_0^2\tau^2}}. \tag{15b}$$

## I. Uncoupled normal diffusion

The spin angular frequency is often affected by a random fluctuating magnetic field, which is produced by surrounding spins undergoing the thermal motion [3,4]; additionally, the angular frequency could be affected by the electron cloud change during the exchange process; further, the chemical exchange may take place because the spin moves among different domains in a heterogeneous system where the frequency fluctuating around positive or negative $\omega_0$. This angular frequency can be denoted as $\omega$, and the average of its absolute value is $\langle|\omega|\rangle = \omega_0$. Because $\omega$ is randomly fluctuating, the individual random phase jump $\Delta\phi = \omega\tau_{jump}$ randomly fluctuates for each jump time $\tau_{jump}$; the space and time uncoupled phase diffusion could be applied to treat the phase random walk with a weak coupling between phase and time distribution; a more complicated coupled diffusion will be considered in the subsequent Section. For an uncoupled diffusion,

$$\langle \tau_{jump}\rangle = \int_0^\infty \frac{t}{\tau}\exp\left(-\frac{t}{\tau}\right)dt = \tau, \tag{16}$$

$$\langle \tau_{jump}^2\rangle = \int_0^\infty \frac{t^2}{\tau}\exp\left(-\frac{t}{\tau}\right)dt = 2\tau^2. \tag{17}$$

The average phase jump length square $\langle(\Delta\phi)^2\rangle$ is

$$\langle(\Delta\phi)^2\rangle = \langle(|\omega|\tau_{jump})^2\rangle = \langle\omega^2\rangle\langle\tau_{jump}^2\rangle = \omega_0^2 2\tau^2 = 2\omega_0^2\tau^2. \tag{18}$$

Such an uncoupled random walk has an effective phase diffusion constant

$$D_\phi = \frac{\langle(\Delta\phi)^2\rangle}{2\langle\tau_{jump}\rangle} = \frac{\langle(\omega\tau_{jump})^2\rangle}{2\tau} = \frac{\langle\omega^2\rangle\langle\tau_{jump}^2\rangle}{2\tau} = \frac{\omega_0^2 2\tau^2}{2\tau} = \omega_0^2\tau. \tag{19}$$

With $D_\phi$, the normal phase diffusion equation can be described by [19,26]

$$\frac{dP(\phi,t)}{dt} = D_\phi \Delta P(\phi,t). \tag{20}$$

From Eq (20), the probability density function is

$$P(\phi,t) = \frac{1}{\sqrt{4\pi D_\phi t}}\exp\left[-\frac{\phi^2}{4D_\phi t}\right]. \tag{21}$$

Substituting Eq. (21) into Eq. (8b), we have

$$p(k,t-t')|_{k=1} = \int_{-\infty}^{\infty}d\phi\, P(\phi,t-t') = \exp[-D_\phi(t-t')], \tag{22}$$

whose Laplace transform representation is

$$p(k,s)|_{k=1} = \frac{1}{s+D_\phi}. \tag{23}$$

Substituting Eqs. (15b) and (23) into Eq. (11) yields

$$M_d(s) = B(s)p(k,s)|_{k=1} = \frac{\frac{1}{1+\omega_0^2\tau^2}}{1+s\frac{\tau(1-\omega_0^2\tau^2)}{1+\omega_0^2\tau^2}}\frac{1}{s+D_\phi}. \tag{24}$$

From $M(s)$, the inverse Laplace transform gives

$$M_d(t) = \int_0^\infty dt\,\frac{1}{\tau(1-\omega_0^2\tau^2)}\exp\left(-\frac{t'}{\frac{\tau(1-\omega_0^2\tau^2)}{1+\omega_0^2\tau^2}}\right)\exp[-D_\phi(t-t')]. \tag{25}$$

Eq. (25) includes the convolution of two parts, the $\frac{1}{\tau(1-\omega_0^2\tau^2)}\exp\left(-\frac{t'}{\frac{\tau(1-\omega_0^2\tau^2)}{1+\omega_0^2\tau^2}}\right)$ comes from the Fourier transform of the $B(t)$ while $\exp[-D_\phi(t-t')]$ results from the phase diffusion; the Frequency domain signal can be obtained from the Fourier Transform of expression (25) as

$$S_d(\omega) = \frac{\frac{1}{1+\omega_0^2\tau^2}}{1+\left[\frac{\tau(1-\omega_0^2\tau^2)}{1+\omega_0^2\tau^2}\right]^2\omega^2} \frac{D_\phi}{D_\phi^2+\omega^2}. \tag{26}$$

II. *Coupled normal diffusion with monoexponential distribution function*

The coupled random phase walk has a joint probability function $\psi(\phi,t)$ expressed by [35,36]

$$\psi(\phi,t) = \varphi(t)\Phi(\phi|t), \tag{27a}$$

$$\Phi(\phi|t) = \frac{1}{2}\delta(|\phi| - \omega_0 t), \tag{27b}$$

where $\varphi(t)$ is the waiting time function, and $\Phi(\phi|t)$ is the conditional probability that a phase jump length $\phi$ requiring time $t$, and $\omega_0 t$ is the absolute value of spin phase change. In Fourier-Laplace representation, the probability density function of a coupled random walk has been derived in Ref. [36] as

$$P(k,s) = \frac{\Psi(k,s)}{1-\psi(k,s)}, \tag{28}$$

where $\Psi(k,s)$ is the Fourier-Laplace representation of the PDF $\Psi(\phi,t)$ for the phase displacement of the last, incomplete walk. $\Psi(\phi,t)$ is defined by

$$\Psi(\phi,t) = \Phi(\phi|t)\Psi(t), \tag{29a}$$

$$\Psi(t) = \int_t^\infty \varphi(t')dt', \tag{29b}$$

where $\Psi(t)$ is the survival probability of random walk [35,36], whose Laplace representation is [35,36]

$$\Psi(s) = \frac{1-\varphi(s)}{s}. \tag{30}$$

For $\Phi(\phi|t) = \frac{1}{2}\delta(|\phi| - \omega_0 t)$, by shift property of Laplace transform, we have

$$\Psi(k,s) = \frac{1}{2}[\Psi(s+ik\omega_0) + \Psi(s-ik\omega_0)], \tag{31}$$

and $$\psi(k,s) = \frac{1}{2}[\varphi(s+ik\omega_0) + \varphi(s-ik\omega_0)]. \tag{32}$$

By substituting $\Psi(k,s)$ and $\psi(k,s)$ into Eq. (28), we get [36]

$$P(k,s) = \frac{\frac{1}{2}[\Psi(s+ik\omega_0)+\Psi(s-ik\omega_0)]}{1-\frac{1}{2}[\varphi(s+ik\omega_0)+\varphi(s-ik\omega_0)]}, \tag{33}$$

which is the same type of equation as the general equation for coupled Levy walk presented in Ref. [36] with the initial condition $P_0(x) = \delta(x)$ and $P_0(k) = 1$. For monoexponential distribution, $\varphi(s) = \frac{1}{\tau s+1}$, based on Eqs. (30-32), we have

$$\Psi(k,s) = \frac{1}{2}\left[\frac{1-\frac{1}{\tau(s+ik\omega_0)+1}}{s+ik\omega_0} + \frac{1-\frac{1}{\tau(s-ik\omega_0)+1}}{s-ik\omega_0}\right] = \frac{1}{2}\left[\frac{\tau}{\tau(s+ik\omega_0)+1} + \frac{\tau}{\tau(s-ik\omega_0)+1}\right] = \frac{\tau(1+\tau s)}{(\tau s+1)^2+k^2\omega_0^2\tau^2}, \tag{34}$$

and

$$\psi(k,s) = \int e^{ik\phi - st}\psi(\phi,t)d\phi dt = \frac{1}{2}\left[\frac{1}{\tau(s-ik\omega_0)+1} + \frac{1}{\tau(s+ik\omega_0)+1}\right] = \frac{1+\tau s}{(\tau s+1)^2+k^2\omega_0^2\tau^2}. \tag{35}$$

Eqs. (34) and (35) can be substituted into Eq. (28) or (33) to give

$$P(k,s) = \frac{\Psi(k,s)}{1-\psi(k,s)} = \frac{\frac{\tau(1+\tau s)}{(\tau s+1)^2+k^2\omega_0^2\tau^2}}{1-\frac{1+\tau s}{(\tau s+1)^2+k^2\omega_0^2\tau^2}}. \tag{36}$$

When k =1,

$$p(k,s)|_{k=1} = \frac{\Psi(1,s)}{1-\psi(1,s)} = \frac{\frac{\tau(1+\tau s)}{(\tau s+1)^2+\omega_0^2\tau^2}}{1-\frac{1+\tau s}{(\tau s+1)^2+\omega_0^2\tau^2}} \approx \frac{\tau(1+\tau s)}{\omega_0^2\tau^2+\tau s}, \quad (37)$$

where the higher-order term $\tau^2 s^2$ is neglected [36]; similar approximate treatments are carried out in this paper. Substituting Eqs. (15b) and (37) into Eq. (11) yields

$$M_d(s) = \frac{1+\tau s}{1+\omega_0^2\tau^2+2\tau s}\frac{\tau(1+\tau s)}{\omega_0^2\tau^2+\tau s} \approx \frac{1}{\tau s(1-\omega_0^2\tau^2)+1+\omega_0^2\tau^2}\frac{\tau}{\tau s(1-\omega_0^2\tau^2)+\omega_0^2\tau^2}$$

$$= \frac{\frac{1}{1+\omega_0^2\tau^2}}{\tau s(1-\omega_0^2\tau^2)/(1+\omega_0^2\tau^2)+1}\frac{\tau/\omega_0^2\tau^2}{\tau s(1-\omega_0^2\tau^2)/\omega_0^2\tau^2+1}. \quad (38)$$

From $M(s)$, the inverse Laplace transform gives

$$M_d(t) = \int_0^t dt' \frac{1}{\tau(1-\omega_0^2\tau^2)}\exp\left(-\frac{t'}{\frac{\tau(1-\omega_0^2\tau^2)}{1+\omega_0^2\tau^2}}\right)\frac{1}{1-\omega_0^2\tau^2}\exp\left(-\frac{t-t'}{\frac{\tau(1-\omega_0^2\tau^2)}{\omega_0^2\tau^2}}\right). \quad (39)$$

Eq. (39) includes the convolution of $\exp\left(-\frac{t'}{\frac{\tau(1-\omega_0^2\tau^2)}{1+\omega_0^2\tau^2}}\right)$ and $\exp\left(-\frac{t-t'}{\frac{\tau(1-\omega_0^2\tau^2)}{\omega_0^2\tau^2}}\right)$, whose Fourier Transform gives the frequency domain NMR signal:

$$S_d(\omega) = \frac{1}{\tau(1-\omega_0^2\tau^2)}\frac{\frac{\tau(1-\omega_0^2\tau^2)}{1+\omega_0^2\tau^2}}{\left(\frac{\tau(1-\omega_0^2\tau^2)}{1+\omega_0^2\tau^2}\right)^2\omega^2+1}\frac{1}{1-\omega_0^2\tau^2}\frac{\frac{\tau(1-\omega_0^2\tau^2)}{\omega_0^2\tau^2}}{\left(\frac{\tau(1-\omega_0^2\tau^2)}{\omega_0^2\tau^2}\right)^2\omega^2+1}$$

$$= \frac{\frac{1}{1+\omega_0^2\tau^2}}{\left(\frac{\tau(1-\omega_0^2\tau^2)}{1+\omega_0^2\tau^2}\right)^2\omega^2+1}\frac{\frac{\tau}{\omega_0^2\tau^2}}{\left(\frac{\tau(1-\omega_0^2\tau^2)}{\omega_0^2\tau^2}\right)^2\omega^2+1}. \quad (40)$$

*2.2.3 Fractional phase diffusion*

For a complicated system, the time correlation function may not be a simple monoexponential function, such as the Kohlrausch-Williams-Watts (KWW) function, or Mittag-Leffler function, and the corresponding phase diffusion could be an anomalous diffusion [30-33].

The time-fractional phase diffusion will be investigated here, and the time correlation function is assumed as an MLF,

$$G(t) = E_\alpha\left(-\left(\frac{t}{\tau}\right)^\alpha\right). \quad (41)$$

Based on Eq. (12b), the frequency domain signal contributed from these residual spins are

$$S_{rs}(\omega) = \frac{1}{2}\left[\frac{(\omega-\omega_0)^{\alpha-1}\tau^\alpha \sin\left(\frac{\pi}{2}\alpha\right)}{(\omega-\omega_0)^{2\alpha}\tau^{2\alpha}+2(\omega-\omega_0)^\alpha\tau^\alpha \cos\left(\frac{\pi}{2}\alpha\right)+1}+\frac{(\omega+\omega_0)^{\alpha-1}\tau^\alpha \sin\left(\frac{\pi}{2}\alpha\right)}{(\omega+\omega_0)^{2\alpha}\tau^{2\alpha}+2(\omega+\omega_0)^\alpha\tau^\alpha \cos\left(\frac{\pi}{2}\alpha\right)+1}\right], \quad (42)$$

The waiting time distribution function will be a heavy-tailed time distribution [37]

$$\varphi_f(t) = -\frac{d}{dt}E_\alpha\left(-\left(\frac{t}{\tau}\right)^\alpha\right), \quad (43a)$$

whose Laplace transform is [35,36]

$$\varphi_f(s) = \frac{1}{s^\alpha\tau^\alpha+1}. \quad (43b)$$

Based on Eqs. (10c), (38) and (39), we have

$$B(s) = Re\varphi_f(s+i\omega_0) = Re\frac{1}{2}\left[\frac{\frac{1}{\tau^\alpha}}{\omega_0^\alpha\left[\cos\left(\frac{\pi}{2}\alpha - \frac{s\alpha}{\omega_0}\right) + i\sin\left(\frac{\pi}{2}\alpha - \frac{s\alpha}{\omega_0}\right)\right] + \frac{1}{\tau^\alpha}}\right]$$

$$\approx \frac{\omega_0^\alpha \tau^\alpha\left(\cos\frac{\pi}{2}\alpha + \frac{1}{\omega_0^\alpha \tau^\alpha}\right)}{1+\omega_0^{2\alpha}\tau^{2\alpha}+2\omega_0^\alpha\tau^\alpha\cos\frac{\pi}{2}\alpha + s\alpha\omega_0^{\alpha-1}\tau^\alpha\sin\frac{\pi}{2}\alpha\frac{1-\omega_0^{2\alpha}\tau^{2\alpha}}{\omega_0^\alpha\tau^\alpha\cos\frac{\pi}{2}\alpha+1}} = \frac{c}{1+s\tau'},$$

(44a)

where

$$c = \frac{\omega_0^\alpha \tau^\alpha\left(\cos\frac{\pi}{2}\alpha + \frac{1}{\omega_0^\alpha \tau^\alpha}\right)}{1+\omega_0^{2\alpha}\tau^{2\alpha}+2\omega_0^\alpha\tau^\alpha\cos\frac{\pi}{2}\alpha},$$

(44b)

and

$$\tau' = \frac{\alpha\omega_0^{\alpha-1}\tau^\alpha\sin\frac{\pi}{2}\alpha\frac{1-\omega_0^{2\alpha}\tau^{2\alpha}}{\omega_0^\alpha\tau^\alpha\cos\frac{\pi}{2}\alpha+1}}{1+\omega_0^{2\alpha}\tau^{2\alpha}+2\omega_0^\alpha\tau^\alpha\cos\frac{\pi}{2}\alpha}.$$

(44c)

*I. Uncoupled fractional diffusion*

For the time-fractional diffusion with MLF-based distribution, the phase diffusion constant can be calculated according to Ref. [19, 31,32] as

$$D_{\phi f} = \frac{\langle(\Delta\phi)^2\rangle}{2\Gamma(1+\alpha)\tau^\alpha}.$$

(45a)

The average phase jump may be assumed as $\langle(\Delta\phi)^2\rangle = \omega_0^2 2\tau^2$, then

$$D_{\phi f} = \frac{\omega_0^2 \tau^{2-\alpha}}{\Gamma(1+\alpha)}.$$

(45b)

With $D_{\phi f}$, the fractional phase diffusion equation can be described by [19,26,30,32,33]

$$_tD_*^\alpha P_f = D_{\phi f}\Delta P_f(\phi,t).$$

(46)

where $0 < \alpha \leq 2$, $D_{f_r}$ is the rotational diffusion coefficient, $a$ is the spherical radius, $_tD_*^\alpha$ is the Caputo fractional derivative defined by [31,32]

$$_tD_*^\alpha f(t) := \begin{cases} \frac{1}{\Gamma(m-\alpha)}\int_0^t \frac{f^{(m)}(\tau)d\tau}{(t-\tau)^{\alpha+1-m}}, & m-1 < \alpha < m, \\ \frac{d^m}{dt^m}f(t), & \alpha = m, \end{cases}$$

Fourier transform of Eq. (46) gives [29,32,33]

$$_tD_*^\alpha p(k,t) = -D_{\phi f}k^2 p(k,t).$$

(47)

The solution of Eq. (47) is $p(k,t) = E_\alpha[-D_{\phi f}k^2 t^\alpha]$ [19,26], whose Laplace representation is [31,32,34]

$$p(k,s) = \frac{s^{\alpha-1}}{s^\alpha + D_{\phi f}k^2},$$

(48)

then

$$p(k,s)|_{k=1} = \frac{s^{\alpha-1}}{s^\alpha + D_{\phi f}}.$$

(49)

Substituting Eqs. (44) and (49) into Eq. (11), we get

$$M_d(s) = \frac{c}{1+s\tau'}\frac{s^{\alpha-1}}{s^\alpha + D_{\phi f}},$$

(50)

whose inverse Laplace transform gives

$$M_d(t) = \int_0^t dt' \frac{c}{\tau'} \exp(-\frac{t'}{\tau'}) E_\alpha[-D_{\phi f}(t-t')^\alpha]. \tag{51}$$

The NMR signal can be obtained from the Fourier transform of M(t), which is

$$S_d(\omega) = B(\omega) \cdot E(\omega) = \frac{c}{1+\tau'^2\omega^2} \cdot \frac{\omega^{\alpha-1}\left(\frac{1}{D_{\phi f}}\right)\sin\left(\frac{\pi}{2}\alpha\right)}{\omega^{2\alpha}\left(\frac{1}{D_{\phi f}}\right)^2 + 2\omega^\alpha\left(\frac{1}{D_{\phi f}}\right)\cos\left(\frac{\pi}{2}\alpha\right) + 1}. \tag{52}$$

*II. Coupled fractional diffusion*

Similar to the coupled normal diffusion presented in Section 2.2, the Fourier-Laplace representation of PDF of time fractional phase diffusion obeys Eq. (33). However, $\varphi_f(s)$ defined by Eq. (43b) is needed to be used for the time-fractional diffusion. Substituted Eq. (43b) into Eq. (32), we have

$$\psi(k,s) = \frac{1}{2}\left[\frac{1}{\tau^\alpha(s-ik\omega_0)^\alpha+1} + \frac{1}{\tau^\alpha(s+ik\omega_0)^\alpha+1}\right]. \tag{53}$$

For $k = 1$,
$$\psi(1,s) \approx \frac{1}{2}\left[\frac{1}{\tau^\alpha(s-i\omega_0)^\alpha+1} + \frac{1}{\tau^\alpha(s+i\omega_0)^\alpha+1}\right] = Re\varphi_f(s+i\omega). \tag{54}$$

Compared to Eq. (44a), it is evident that

$$\psi(1,s) = B(s) = \frac{c}{1+s\tau'}. \tag{55}$$

Additionally, it needs to use the $\varphi_f(s)$ in Eq. (43) to calculate the Fourier-Laplace representation of the joint survival probability $\Psi(1,s)$, $k=1$. Based on Eqs. (30), (31) and (43b), we have

$$\Psi(1,s) = \frac{1}{2}\left[\frac{1-\frac{1}{(s+i\omega_0)^\alpha\tau^\alpha+1}}{s+i\omega_0} + \frac{1-\frac{1}{(s-i\omega_0)^\alpha\tau^\alpha+1}}{s-i\omega_0}\right] = \frac{c_1}{1+s\tau'_1}, \tag{56a}$$

where

$$c_1 = \frac{\tau^\alpha \omega_0^{\alpha-1}\sin\frac{\pi}{2}\alpha}{\omega_0^{2\alpha}\tau^{2\alpha} + 2\omega_0^\alpha\tau^\alpha\cos\frac{\pi}{2}\alpha + 1}, \tag{56b}$$

$$\tau'_1 = \frac{2\alpha\omega_0^{\alpha-1}\tau^\alpha\sin\frac{\pi}{2}\alpha}{\left(\omega_0^{2\alpha}\tau^{2\alpha} + 2\omega_0^\alpha\tau^\alpha\cos\frac{\pi}{2}\alpha + 1\right)} - \frac{\omega_0^\alpha\tau^\alpha - (\alpha-1)\cos\left(\frac{\pi}{2}\alpha\right)}{\omega_0\sin\frac{\pi}{2}\alpha}. \tag{56c}$$

Eqs. (55) and (56) can be substituted into Eq. (33) to give

$$p(k,s)|_{k=1} = \frac{\Psi(1,s)}{1-\psi(1,s)} = \frac{\frac{c_1}{1+s\tau'_1}}{1-\frac{c}{1+s\tau'}}. \tag{57}$$

By substituting Eqs. (44a) and (57) into Eq. (11), we get

$$M_d(s) = B(s)p(k,s)|_{k=1} = \frac{c}{1+s\tau'}\frac{\frac{c_1}{1+s\tau'_1}}{1-\frac{c}{1+s\tau'}} = \frac{\frac{c}{1-c}}{1+s\frac{\tau'}{1-c}}\frac{c_1}{1+s\tau'_1}, \tag{58}$$

whose inverse Laplace transform yields

$$M_d(t) = \int_0^t dt' \frac{c}{\tau'}\exp\left(-\frac{t}{\frac{\tau'}{1-c}}\right)\frac{c_1}{\tau'_1}\exp\left(-\frac{t-t'}{\tau'_1}\right). \tag{59}$$

Eq. (59) involves the convolution of two exponential functions, and its Fourier transform can be obtained, which is the NMR frequency domain signal

$$S_d(\omega) = \frac{c}{\tau'}\frac{\frac{\tau'}{1-c}}{1+\left(\frac{\tau'}{1-c}\right)^2\omega^2}\frac{c_1}{\tau'_1}\frac{\tau'_1}{1+\tau'^2_1\omega^2} = \frac{\frac{c}{1-c}}{1+\left(\frac{\tau'}{1-c}\right)^2\omega^2}\frac{c_1}{1+\tau'^2_1\omega^2}. \tag{60}$$

## 3. Results

A phase diffusion equation method is proposed to describe the effect of chemical exchange on the NMR spectrum, based on uncoupled and coupled normal and fractional diffusions. The exchange between two sites with equal populations is considered, and the theoretical expressions are organized in Table 1.

**Table 1**

Comparison of theoretical NMR line shape expressions from phase diffusion method to traditional results for chemical exchange between two sites with equal populations.

| Frequency domain signal expression from phase diffusion results: |
|---|
| Simple phase diffusion with a fixed jump (FJ) time |
| $S(\omega) = \frac{D_{\phi,s}}{D_{\phi,s}^2+\omega^2}$, $D_{\phi,s} = \frac{\omega_0^2}{2}\tau$. |
| Normal phase diffusion with monoexponential function $S(\omega) = S_d(\omega) + S_{rs}(\omega)$, |
| $S_{rs}(\omega) = \frac{1}{2}\left[\frac{\tau}{[\tau(\omega-\omega_0)]^2+1} + \frac{\tau}{[\tau(\omega+\omega_0)]^2+1}\right]$. |
| Uncoupled normal (UN) diffusion |
| $S_d(\omega) = \frac{\frac{1}{1+\omega_0^2\tau^2}}{1+\left[\frac{\tau(1-\omega_0^2\tau^2)}{1+\omega_0^2\tau^2}\right]^2\omega^2}\frac{D_\phi}{D_\phi^2+\omega^2}$, $D_\phi = \omega_0^2\tau$. |
| Coupled normal (CN) diffusion |
| $S_d(\omega) = \frac{\frac{1}{1+\omega_0^2\tau^2}}{\left(\frac{\tau(1-\omega_0^2\tau^2)}{1+\omega_0^2\tau^2}\right)^2\omega^2+1}\frac{\frac{\tau}{\omega_0^2\tau^2}}{\left(\frac{\tau(1-\omega_0^2\tau^2)}{\omega_0^2\tau^2}\right)^2\omega^2+1}$. |
| Fractional phase diffusion with heavy-tailed time distribution $S(\omega) = S_d(\omega) + S_{rs}(\omega)$, |
| $S_{rs}(\omega) = \frac{1}{2}\left[\frac{(|\omega-\omega_0|)^{\alpha-1}\tau^\alpha\sin\left(\frac{\pi}{2}\alpha\right)}{(|\omega-\omega_0|)^{2\alpha}\tau^{2\alpha}+2(|\omega-\omega_0|)^\alpha\tau^\alpha\cos\left(\frac{\pi}{2}\alpha\right)+1} + \frac{(|\omega+\omega_0|)^{\alpha-1}\tau^\alpha\sin\left(\frac{\pi}{2}\alpha\right)}{(|\omega+\omega_0|)^{2\alpha}\tau^{2\alpha}+2(|\omega+\omega_0|)^\alpha\tau^\alpha\cos\left(\frac{\pi}{2}\alpha\right)+1}\right]$, |
| $c = \frac{\omega_0^2\tau^\alpha\left(\cos\frac{\pi}{2}\alpha+\frac{1}{\omega_0^\alpha\tau^\alpha}\right)}{1+\omega_0^{2\alpha}\tau^{2\alpha}+2\omega_0^\alpha\tau^\alpha\cos\frac{\pi}{2}\alpha}$, $\tau' = \frac{\alpha\omega_0^{\alpha-1}\tau^\alpha\sin\frac{\pi}{2}\alpha\frac{1-\omega_0^{2\alpha}\tau^{2\alpha}}{\omega_0^\alpha\tau^\alpha\cos\frac{\pi}{2}\alpha+1}}{1+\omega_0^{2\alpha}\tau^{2\alpha}+2\omega_0^\alpha\tau^\alpha\cos\frac{\pi}{2}\alpha}$. |
| Uncoupled fractional (UF) diffusion |
| $S_d(\omega) = \frac{c}{1+\tau'^2\omega^2}\cdot\frac{|\omega|^{\alpha-1}\left(\frac{1}{D_{\phi f}}\right)\sin\left(\frac{\pi}{2}\alpha\right)}{|\omega|^{2\alpha}\left(\frac{1}{D_{\phi f}}\right)^2+2|\omega|^\alpha\left(\frac{1}{D_{\phi f}}\right)\cos\left(\frac{\pi}{2}\alpha\right)+1}$, $D_\phi = \frac{\omega_0^2\tau^2}{\Gamma(1+\alpha)\tau^\alpha}$. |
| Coupled fractional (CF) diffusion |
| $S_d(\omega) = \frac{\frac{c}{1-c}}{1+\left(\frac{\tau'}{1-c}\right)^2\omega^2}\frac{c_1}{1+\tau'^2_1\omega^2}$, |
| where $c_1$ and $\tau'_1$ are defined in Eqs. (56b) and (56c), respectively. |
| Frequency domain signal expression from the traditional (TR) method: |
| $S(\omega) = \frac{\omega_0^2\frac{\tau}{2}}{\left[\frac{\tau}{2}(\omega_0^2-\omega^2)\right]^2+\omega^2}$ [1,2,8]. |

## 4. Discussion

In the rotating frame reference, the spin phase in chemical exchange undergoes random phase jumps, which can be intrinsically described by either an uncoupled effective phase diffusion equation or coupled random walk.

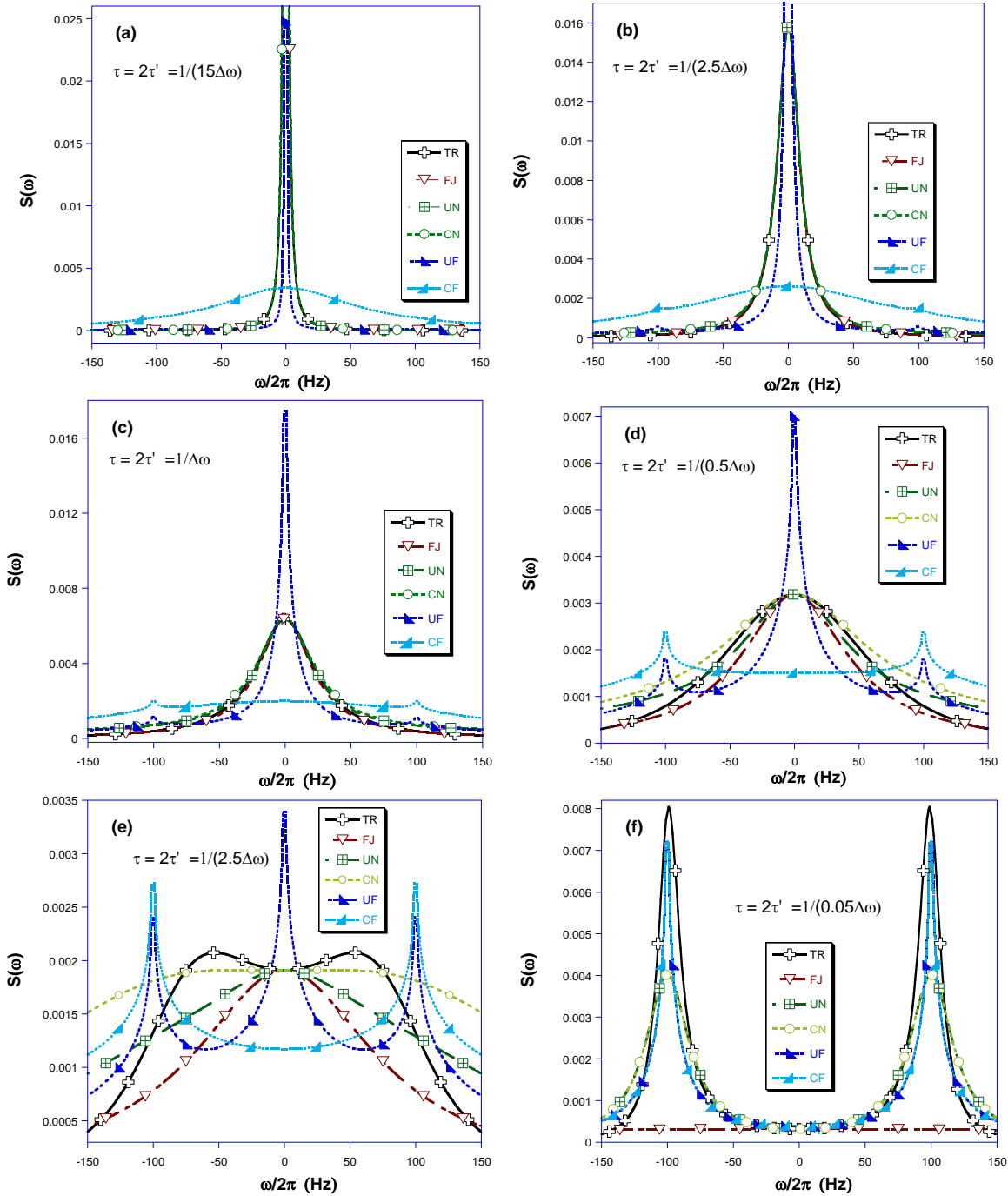

**Fig. 1** The comparison among the theoretical line shapes $S(\omega)$ with various exchange times. $\tau = 2\tau'$, and $\tau$ equals $\frac{1}{15\Delta\omega}, \frac{1}{2.5\Delta\omega}, \frac{1}{\Delta\omega}, \frac{1}{0.5\Delta\omega}, \frac{1}{0.3\Delta\omega}$, and $\frac{1}{0.05\Delta\omega}$, respectively. The employed models are listed in Table 1, including traditional (TR), fixed jump (FJ) time, uncoupled normal (UN) diffusion, coupled normal (CN) diffusion, uncoupled fractional (UF) diffusion, and coupled fractional (CF) diffusion models. $\tau$ is used for both the TR and FJ models, while $\tau'$ is for UN, CN, UF, and CF models. Additionally, $\Delta\omega = 2\omega_0$, $\omega_0/2\pi = 100$ Hz, and $\alpha = 0.75$ is used for all fractional diffusions.

Figure 1 compares the theoretical results with different exchange times obtained from the phase diffusion models and those obtained by the traditional two-site exchange model. All equations used are listed in Table 1. $\tau$ is the exchange time constant for the traditional model and the fixed time diffusion, while $\tau'$ is for the coupled and uncoupled normal diffusion and fractional diffusion. In Figures 1a-f, $\tau$ equals $2\tau'$, and $\tau$ are set as $\frac{1}{15\Delta\omega}, \frac{1}{2.5\Delta\omega}, \frac{1}{\Delta\omega}, \frac{1}{0.5\Delta\omega}, \frac{1}{0.3\Delta\omega}$, and $\frac{1}{0.03\Delta\omega}$, respectively. $\Delta\omega/2\pi = 100$ Hz. From Figure 1, when the exchange is sufficiently fast in (a-c), $\tau = 2\tau' \leq 1/\Delta\omega$, the theoretical curves from diffusion with a fixed jump time, uncoupled and coupled normal diffusion almost overlap with that predicted from the traditional model; however, the exchange time constant $\tau$ for the traditional model and the fixed time diffusion is two times as $\tau'$ for the coupled and uncoupled normal diffusion with the monoexponential distribution. The difference in exchange time could be explained by the following: the effective phase diffusion constant is $\frac{\omega_0^2}{2}\tau$ for diffusion with a fixed jump time $\tau$; in contrast, it is $\omega_0^2\tau$ for the uncoupled diffusion with a monoexponential time distribution. The two-time difference in diffusion coefficients resulted from the $\langle \tau_{jump}^2 \rangle = \int_0^\infty \frac{t^2}{\tau} \exp\left(-\frac{t}{\tau}\right) dt = 2\tau^2$, while in the fixed time jump, $\langle \tau_{jump}^2 \rangle = \tau^2$. The same phase diffusion coefficient $\omega_0^2\tau$ based on monoexponential time distribution function has been used in Ref. [26] to obtain NMR relaxation expressions, which replicate the traditional NMR relaxation theories; these traditional NMR relaxation expressions have been used to explain many experimental results [1,2,38]; although this theoretical and experimental confirmation is from relaxation NMR, it still provides solid support to select $\omega_0^2\tau$ rather than $\frac{\omega_0^2}{2}\tau$ as a phase diffusion coefficient, considering both the exchange and relaxation are random phase walk processes. Therefore, in the NMR chemical exchange line shape analysis, the exchange time constant could have a two-time difference depending on the employed models.

However, in Figures 1e-f, the theoretical curves from the three normal phase diffusion models: UN, CN, and FJ deviate from the curve based on the traditional model. This deviation is a challenge for the diffusion model. At the slow exchange, the spin phase random walk should not have enough time to meet the diffusion limit; therefore, the possible phase distribution in the system should not obey the probability function obtained from a diffusion equation. A suitable phase distribution function is needed to give a good line shape for the slow exchange. However, for fractional phase diffusion, the applicable range could be different.

Additionally, in Figure 1, the exchange line shapes in normal and fractional diffusion are significantly different. The spectrum line from coupled fractional diffusion is broader than that of uncoupled fractional diffusion, which may be reasonable because heavy-tailed time distribution has a more direct effect on the phase length in the coupled fractional diffusion than that of uncoupled fractional diffusion. Meanwhile, the difference between the coupled and uncoupled diffusion is negligible in normal diffusion but significant in fractional diffusion; compared to the peak from the traditional model, the peak from the coupled fractional diffusion is broader, while it is narrower from the uncoupled fractional diffusion.

Furthermore, from figures 1a-f, when $\tau'$ increases, in uncoupled fractional phase diffusion with a heavy-tailed time distribution, the signal peak changes from a signal peak to a triple peak, then a double peak, while in the normal phase diffusion with monoexponential time distribution, the single peak changes to a double peak directly, no triple peak appearing. The curve of the coupled fractional phase diffusion in Figure 1c shows an ambiguous triple peak because the middle peak is broad, and the two side peaks are small and narrow. The traditional method could interpret the triple peak pattern from a heavy-tailed time distribution as a binodal exchange. The origin of the triple peak can be seen more evident in Figures 2a-f.

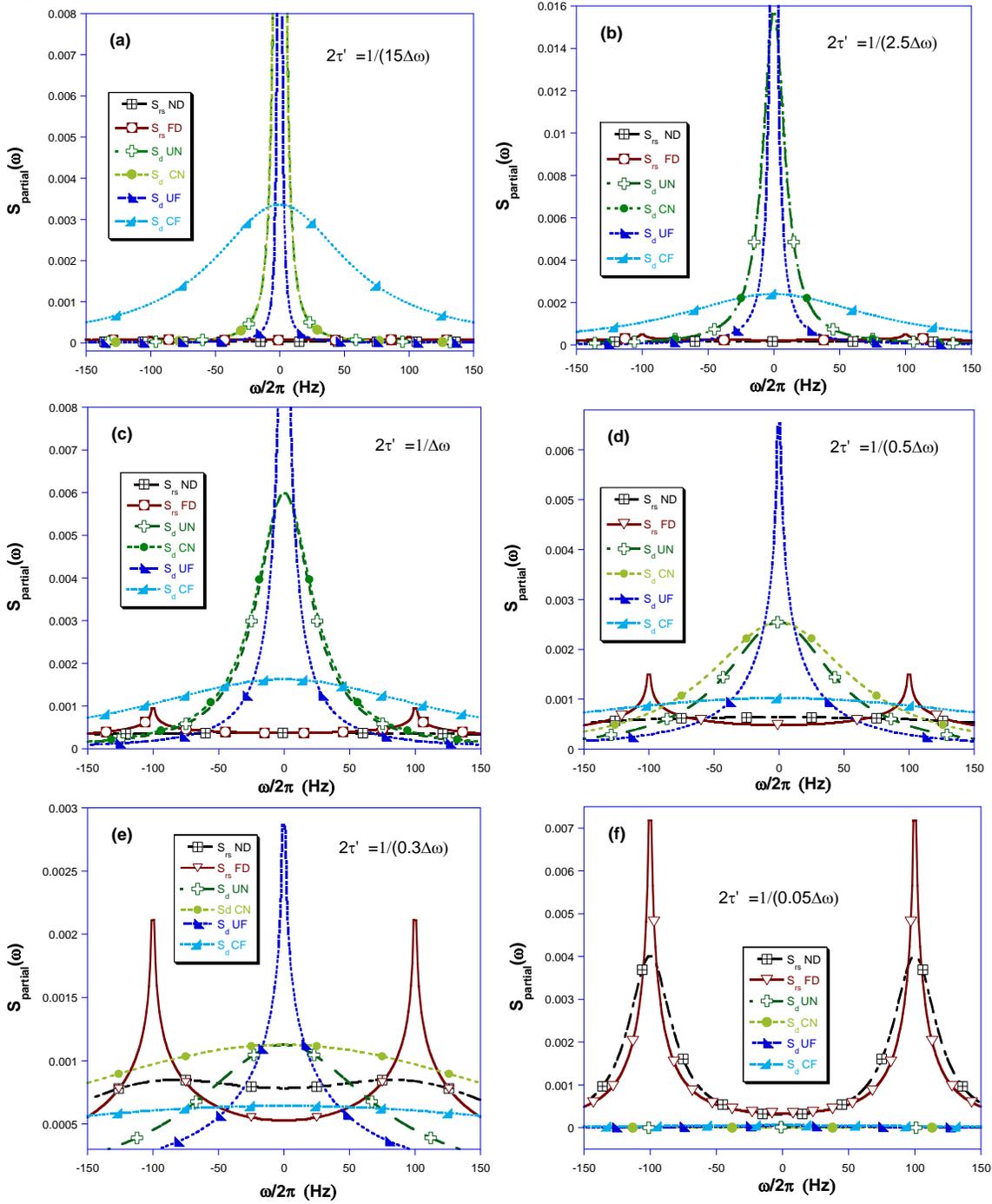

**Fig. 2** The comparison among the partial spectrum intensities $S_{partial}(\omega)$ from the diffusing spins and the residual spins, namely $S_d(\omega)$ and $S_{rs}(\omega)$, at various exchange times, $\tau'$. $2\tau'$ equals $\frac{1}{15\Delta\omega}$, $\frac{1}{2.5\Delta\omega}$, $\frac{1}{\Delta\omega}$, $\frac{1}{0.5\Delta\omega}$, $\frac{1}{0.3\Delta\omega}$, and $\frac{1}{0.05\Delta\omega}$, respectively. Four diffusion models, UN, CN, UF and CF, listed in Table 1 are predicted. Other parameters used in the calculation are $\Delta\omega = 2\omega_0$, $\omega_0/2\pi = 100$ Hz, and both the coupled and uncoupled fractional phase diffusions have $\alpha = 0.75$.

Figures 2a-f compare the frequent domain signal intensities stemming from the diffusing spins and residual spins, $S_d(\omega)$ and $S_{rs}(\omega)$ defined by Eqs. (12b) and (12c), respectively. In NMR exchange experiments, the signal of spins magnetization at time *t* comes from both the spins that have diffused and those residual spins that have not diffused yet. The $S_d(\omega)$ and $S_{rs}(\omega)$ dominate the signal intensity at the fast exchange and the slow exchange, respectively. When the exchange becomes slower, the single peak from the diffusing spins becomes broader, and its height decreases, and eventually, the single peak becomes negligible at the slow exchange, while, the residual peak $S_{rs}(\omega)$ is always a double peak, but it is very small and negligible at the fast exchange. The residual double-peak increases faster in the fractional exchange than in the normal exchange when exchange becomes slower; therefore, a triple peak is observed in the total signal intensity $S(\omega)$ shown in Figures 1b-e. The triple-peak is not observed in the normal phase diffusion model because the residual double-peak shows up after the diffusing peak has become too broad and small.

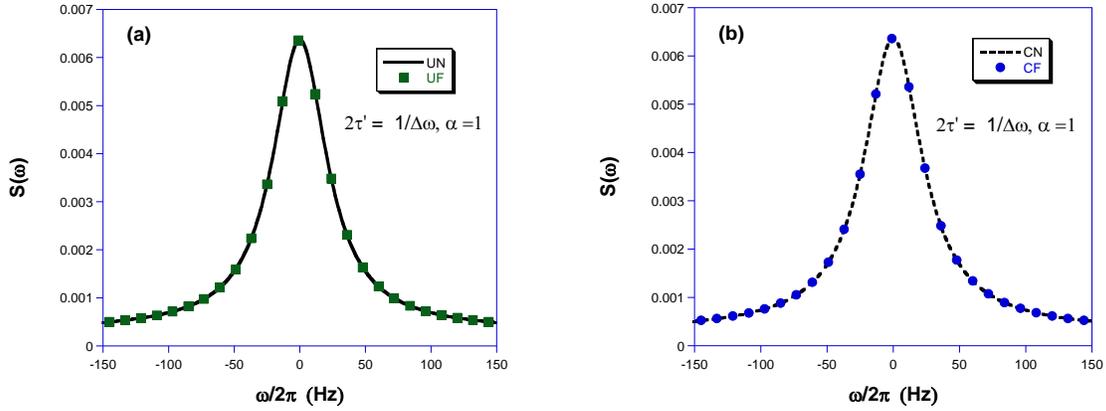

**Fig. 3** The fractional diffusion results reduce to the normal diffusion results when fractional derivative order $\alpha$ equals 1. Other parameters used are $\Delta\omega = 2\,\omega_0$, $\omega_0/2\pi = 100$ Hz, and $2\tau' = 1/\Delta\omega$. In both (a) uncoupled and (b) coupled diffusions, the curves from normal and fractional diffusion with $\alpha = 1$ are overlapped.

In Figure 3a for uncoupled diffusion and Figure 3b for coupled diffusion, the curves from both normal diffusion and fractional diffusion with $\alpha = 1$ are overlapped, which implies that the fractional diffusion results reduce to the normal diffusion results when $\alpha = 1$. This overlap is reasonable because normal diffusion can be viewed as a specific case of fractional diffusion. Other parameters used in the calculation of the curves in Figures 2a-f are $\Delta\omega/2\pi = 100$ Hz, and $2\tau' = 1/\Delta\omega$.

In Figure 4, the theoretical curves from the (a) uncoupled and (b) coupled fractional phase diffusion show different trends when the time-fractional derivative parameter $\alpha$ changes. $\alpha$ = 1, 0.9, 0.75, and 0.5 are used in Figures 4a-b. When $\alpha$ decreases, in the uncoupled fractional diffusion, the signal peak becomes narrower and eventually to a triple peak at $\alpha$ =0.5, while in the coupled fractional diffusion, NMR peak becomes broader and becomes a double peak rather than a triple peak at $\alpha$ =0.5. Additionally, the line shapes differ between the uncoupled and coupled fractional diffusion curves. The line shape of coupled fractional diffusion looks like a bell shape, while the middle peak of the uncoupled fractional exchange looks like a Lorentzian shape. In the view of the traditional model, this line shape of uncoupled fractional exchange could be interpreted as a bimodal exchange. However, both the fast and slow exchange times come from the same heavy-tailed time distribution.

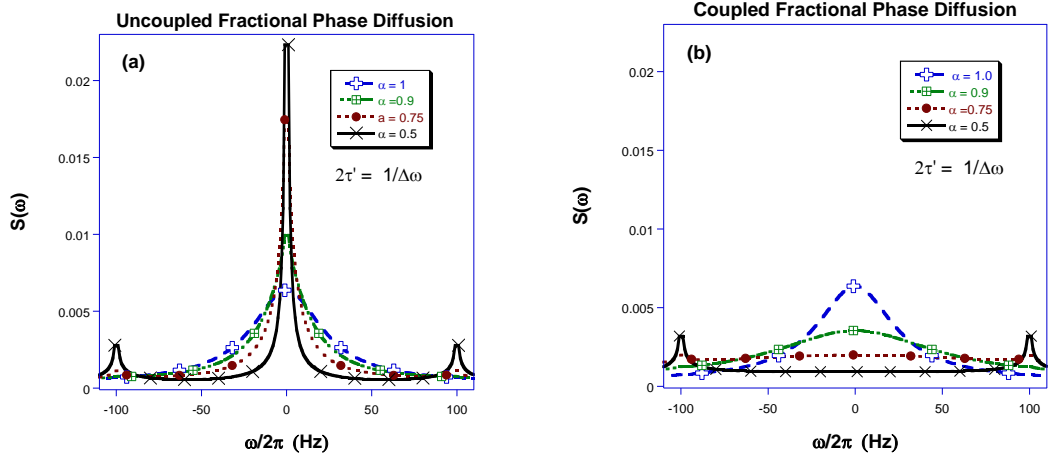

**Fig. 4** The changes of line shapes among different fractional derivative orders, $\alpha$ = 1, 0.9, 0.75 and 0.5 in (a) uncoupled fractional diffusion and (b) coupled fractional diffusions. Other parameters used are $\Delta\omega = 2\omega_0$, $\omega_0/2\pi = 100$ Hz, and $2\tau' = 1/\Delta\omega$.

In Figures 2 and 4, for the uncoupled fractional diffusion, these $S_d(\omega)$ peaks depend on the fractional diffusion coefficient defined by Eq. (45b), $D_{\phi f} = \frac{\omega_0^2 \tau^{2-\alpha}}{\Gamma(1+\alpha)}$. Such a fractional diffusion coefficient is approximately adopted because the second moment of the jump length is divergence. This difficulty of such a divergence is remedied by the coupled fractional diffusion [36] because each jump length is acquired with a time cost. Researchers may adopt a different approximate uncoupled fractional diffusion coefficient, which, however, should be able to reduce to the diffusion coefficient of normal diffusion when $\alpha$ equals 1. A different diffusion coefficient gives a different $S_d(\omega)$ peak, which affects the height of the uncoupled fractional diffusion and the relative ratio between the $S_d(\omega)$ and the $S_{rs}(\omega)$ peaks.

The theoretical expression of normal phase diffusion expression could be applied to small molecule systems, where the phase diffusion coefficient $D_\phi$ could be treated as linewidth when other factors, such as relaxation, are neglected. In contrast, the expressions from fractional diffusion could be applied to polymer or biological systems. It still needs more effort to obtain the relationship between the linewidth and diffusion parameters, such as exchange time, and derivative order. Different models have distinct line shapes, determining how to choose the desired model.

In the derivation, the NMR relaxation is neglected. The relaxation effect changes the linewidth of the NMR peaks. In general, a spin-spin relaxation time $T_2$ yields a linewidth $\frac{1}{T_2}$ (equal to relaxation rate). When the relaxation cannot be neglected, the line shape is the combined result of exchange and relaxation. At the fast exchange, the total linewidth is $\frac{1}{T_2} + D_\phi$. To obtain the characteristic exchange time based $\tau = D_\phi/\omega_0^2$ from the linewidth, the $\frac{1}{T_2}$ effect on linewidth needs to be considered. Additionally, relaxation could have a different effect on $S_d(\omega)$ and $S_{rs}(\omega)$. In NMR relaxation, fast-moving spins and slow spins often have different relaxation rates. The signal from residual spins could decay faster than that from these diffusing spins and get more broadening from the spin-spin relaxation. The $S_{rs}(\omega)$ are affected by $T_2$, while $S_d(\omega)$ is affected by phase random walk from both the exchange and relaxation. Because both relaxation and exchange can be described by phased diffusion [26], the current model could be extended to handle the relaxation and exchange together to give a general expression, which is the future research in this direction.

In this method, the Laplace and inverse Laplace transforms are used in the theoretical derivation, which does not involve numerical computing in obtaining the NMR exchange spectrum. Fourier transform is often employed to transfer the NMR time domain signal to the frequency domain signal in practical NMR experiments, which is not affected by the difficulty of the ill-posed nature of Inverse Laplace Transformation in the numerical calculation. Nevertheless, if the methodology is extended to treat other exchange phenomena, where the numerical implementation of inverse Laplace transform could be needed, the ill-posed issues may appear and need to be handled by certain regularization and stabilization methods.

Besides the line shape change, other dynamic information, such as diffusion and relaxation, has been employed to investigate exchange [5-7,39]. The methodology here could be adapted to approximately handle a random exchange process in a system where the exchange sites do not have spectroscopic separation. For instance, if the exchangeable sites have different decay (such as signal attenuation), the absolute average decay can be used as a reference decay. The fast and slow decay will be assigned as relative loss and gain, respectively. This random relative gain and loss process could be treated as a random walk and described by an effective diffusion equation. However, unlike the phase diffusion coefficient in this paper, the diffusion coefficient for such a gain-loss random walk process should be time-dependent because the absolute average decay decreases with time. If an appropriate diffusion coefficient can be derived, the distribution of possible loss or gain (relative to the reference decay) could be obtained, which may provide meaningful information to the application. Of course, the actual situation can be different and challenging, and these applications could involve inverse Laplace transformations in analyzing NMR data. Extending this method to these applications needs a detailed derivation, which may not be obtainable.

Both the traditional method and the current phase diffusion method have limits in interpreting the NMR chemical exchange phenomena. For the traditional model, it is often derived based on tedious exchange matrix calculation with a single fixed exchange time constant, and the detailed results are often available only for two or three sites exchange; additionally, it is challenging to handle exchange time with a distribution and fractional exchange. While for the current phase diffusion method, it encounters challenges in slow exchange, although it shows excellent results in the fast exchange range. The difficulty in slow exchange results from the fact that the diffusion limit may not be met because the experimental time window in NMR is not infinite. Additionally, it neglects the relaxation effect and considers only the fundamental model, the two sites exchange with a symmetric population. However, the current phase diffusion method is still in its early stages; with more future effort in this direction, these limits could be overcome. The current method can be applied based on other anomalous diffusion models, such as the fractal derivative [40,41,42]. Further research is needed to understand and apply the method proposed in this paper, particularly the fractional diffusion model, and to extend the current method for multiple sites and unequal population exchange.

5. **Conclusion**

This paper proposes a phase diffusion method to describe the chemical exchange NMR spectrum. This method directly analyzes the spin system evolution in phase space, which is convenient for handling complex exchanges with fractional calculus. The method is applied to study the fundamental exchange, a two-site exchange with each population. Currently, the method can handle fast to medium exchange but still has difficulty in slow exchange. The theoretical results show significant differences between normal and fractional exchange. First, the line shape difference between coupled and uncoupled diffusion is not evident in normal diffusion but significant in fractional diffusion; second, there is a significant difference in the line shape pattern: When the exchange becomes slower, in the uncoupled fractional phase diffusion, the single peak becomes a triple peak first, then two peaks eventually, while in the normal exchange, the single peak becomes two peaks directly. Unlike the traditional method, the exchange time constant can follow certain types of distributions. Additionally, the exchange time constant is two times faster based on the monoexponential time distribution than that obtained by the traditional model. Furthermore, this phase diffusion method could be combined with other phase diffusion equations in relaxation and PFG diffusion to deal with more complicated scenarios.